\documentclass[%
 aip,
 jmp,%
 amsmath,amssymb,
preprint,%
]{revtex4-1}

\usepackage{graphicx}
\usepackage{dcolumn}
\usepackage{bm}

\makeatletter 
\newcommand*{\citenst}[2][]{%
  \begingroup
  \let\NAT@mbox=\mbox
  \let\@cite\NAT@citenum
  \let\NAT@space\NAT@spacechar
  \let\NAT@super@kern\relax
  \renewcommand\NAT@open{}%
  \renewcommand\NAT@close{}%
  \cite[#1]{#2}%
  \endgroup
}
\makeatother

\usepackage[colorlinks]{hyperref}  

\begin{document}


\title[Quantification of the spin-Hall anti-damping torque with a resonance spectrometer]{Quantification of the spin-Hall anti-damping torque with a resonance spectrometer}

\author{Satoru Emori}%
\email{
s.emori@neu.edu
}
\affiliation{ 
Department of Electrical and Computer Engineering, Northeastern University, Boston, MA 02115
}%
\author{Tianxiang Nan}%
\affiliation{ 
Department of Electrical and Computer Engineering, Northeastern University, Boston, MA 02115
}%
\author{Trevor M. Oxholm}%
\affiliation{ 
Department of Electrical and Computer Engineering, Northeastern University, Boston, MA 02115
}%
\author{Carl T. Boone}%
\affiliation{ 
Department of Electrical and Computer Engineering, Northeastern University, Boston, MA 02115
}%
\author{John G. Jones}%
\affiliation{ 
Materials and Manufacturing Directorate, Air Force Research Laboratory, Wright-Patterson AFB, OH 45433
}%
\author{Brandon M. Howe}%
\affiliation{ 
Materials and Manufacturing Directorate, Air Force Research Laboratory, Wright-Patterson AFB, OH 45433
}%
\author{Gail J. Brown}%
\affiliation{ 
Materials and Manufacturing Directorate, Air Force Research Laboratory, Wright-Patterson AFB, OH 45433
}%
\author{David E. Budil}%
\affiliation{ 
Department of Chemistry, Northeastern University, Boston, MA 02115
}%
\author{Nian X. Sun}%
\affiliation{ 
Department of Electrical and Computer Engineering, Northeastern University, Boston, MA 02115
}%

\begin{abstract}
We present a simple technique using a cavity-based resonance spectrometer to quantify the anti-damping  torque due to the spin Hall effect. 
Modification of ferromagnetic resonance is observed as a function of small DC current in sub-mm-wide strips of bilayers, consisting of magnetically soft FeGaB and strong spin-Hall metal Ta.  
From the detected current-induced linewidth change, we obtain an effective spin Hall angle of 0.08-0.09 independent of the magnetic layer thickness.   
Our results demonstrate that a sensitive resonance spectrometer can be a general tool to investigate spin Hall effects in various material systems, even those with vanishingly low conductivity and magnetoresistance.  
\end{abstract}

\maketitle

Efficient and scalable spintronic memory, logic, and signal generation devices require direct manipulation of magnetic moments by electric current.~\cite{Brataas2012, Locatelli2014}   
Current-induced torques arising from spin-orbit phenomena~\cite{Brataas2014, Haney2013a} have recently emerged as a robust means of controlling magnetization in ferromagnet/normal-metal bilayer thin films.  
In particular, the spin Hall effect in the normal metal converts an in-plane charge current to an out-of-plane spin current,~\cite{Hoffmann2013} which exerts a torque on magnetic moments in the adjacent ferromagnetic layer.
This spin-Hall torque can counteract damping in the ferromagnet and has been shown to switch uniform magnetization,~\cite{Miron2011, Liu2012, Liu2012a, Pai2012} drive domain walls,~\cite{Haazen2013, Emori2013, Emori2013a} and control precessional magnetization dynamics.~\cite{Ando2008a, Demidov2011, Liu2011, Pai2012, Wang2011, Wang2011a, Liu2012, Demidov2012, Liu2013, Zholud2014, Ganguly2014, Kasai2014, Skinner2014}   

Despite the demonstrated utility of the spin Hall effect, there exists a wide disparity in its reported magnitude parameterized by the spin Hall angle.
For example, reported spin Hall angles for Ta, one of the most commonly studied spin-Hall metals, span more than an order of magnitude based on various techniques.~\cite{Morota2011, Liu2012, Deorani2013,Zhang2013a, Emori2013a, Fan2014, Hahn2013, Wang2014, Qu2014}
Because the spin Hall effect generates an anti-damping torque (Fig. 1), a conceptually straightforward technique is to measure the change in magnetization damping as a function of DC charge current.  
This has recently been done extensively in micron-wide strips of NiFe/Pt bilayers using spin-torque ferromagnetic resonance (ST-FMR).~\cite{Liu2011, Ganguly2014, Kasai2014}  
However, it is challenging to resolve DC-current-induced linewidth changes in ST-FMR spectra in material systems with high resistivity or low anisotropic magnetoresistance.  

Preceding the recent ST-FMR studies by a few years, Ando \emph{et al.} measured current-induced linewidth modifications in NiFe/Pt with a cavity-based resonance spectrometer.~\cite{Ando2008a}  
This technique is the inverse of measuring the DC voltage from spin pumping in a spin-Hall bi(multi)layer placed inside a microwave cavity.~\cite{Saitoh2006, Ando2011, Heinrich2011, Azevedo2011, Weiler2013, Wang2013, Wang2014}  
Owing to its high sensitivity, a cavity-based spectrometer may also be an excellent tool for detecting the spin-Hall anti-damping torque, although this technique has not been exploited for systematic quantification of spin Hall effects.  

In this Letter, we demonstrate a simple, general method using a cavity-based resonance spectrometer for quantifying the spin-Hall anti-damping torque and effective spin Hall angle in ferromagnet/normal-metal bilayers.
The bilayers are patterned into sub-mm wide strips, narrow enough to attain charge current densities of $\sim$10$^9$ A/m$^2$, yet large enough to attain a signal-to-noise ratio that permits resolving small linewidth changes of $\sim$10 $\mu$T.     
Magnitude of the anti-damping torque scales inversely with the thickness of the magnetic layer, and we obtain an effective spin Hall angle independent of the magnetic layer thickness.  
These results are consistent with the spin-Hall mechanism of the anti-damping torque and confirm that a cavity-based spectrometer is a viable tool for reliable quantification of spin Hall effects.  

We select a magnetically soft alloy of Fe$_7$Ga$_2$B$_1$ (herein denoted as FeGaB) as the ferromagnetic material because of its narrow resonance linewidth~\cite{Lou2007} that enables precise detection of small current-induced resonance modifications.  
Moreover, the resistivities of nanometer-thick FeGaB and Ta are nearly identical at $\approx$200 $\mu \Omega$cm, permitting a convenient assumption of uniform charge current distribution in the bilayer.  
The high resistivity, compounded by the low anisotropic magnetoresistance, of FeGaB/Ta also makes it difficult for measurement with ST-FMR, which brings out the cavity-based method as a reliable alternative with a higher signal-to-noise ratio.  

FeGaB/Ta strips, 100 and 200 $\mu$m wide and 1.5 mm long, were patterned using photolithography, magnetron sputter deposition, and liftoff.  The substrate was Si(001) with 50 nm of thermally-grown SiO$_2$.  
The FeGaB layer was co-sputtered from Fe$_{8}$Ga$_{2}$ (DC sputtered) and B (RF sputtered) targets.  
Samples with nominal FeGaB thicknesses $t_{F}$ = 2.2, 2.9, 3.4, and 4.4 nm were prepared.  
The Ta overlayer was DC sputtered, and its thickness $t_{Ta}$ was 5.5 nm for all samples with 2 nm of passivated oxide layer on top.  
The uncertainty in $t_{F}$ and $t_{Ta}$ is estimated to be $<$10\% from x-ray reflectivity.  
We note that $t_{Ta}$ here is significantly greater than the typical reported spin diffusion length in Ta of $\lambda_{Ta}\approx$1 nm.~\cite{Zhang2013a, Qu2014, Boone2013}  
Cr(5 nm)/Au(120 nm) electrodes, with an overlap length of 25 $\mu$m at each end of the FeGaB/Ta strip, were also defined by photolithography and deposited with DC sputtering.  

FMR measurements were conducted using a Bruker EMX electron paramagnetic resonance spectrometer with a TE102 cavity, operated at a microwave field frequency of 9.76 GHz and power of 10 mW.  
The sample was placed at the center of the cavity, which was optimized such that its tuning was not affected by varying the DC current.   
A Keithley 6220 current source was used to output the DC current through the strip. The typical device resistance, including contact resistance, was $\approx$2-4  k$\Omega$. 
The power dissipation through the device was limited to $\ll$100 mW, i.e., $|I_{DC}| \ll 10$ mA, to avoid irreversible annealing effects due to Joule heating.  

Figure 2(a) defines the essential parameters defining the ferromagnetic resonance (FMR) spectrum: resonance field $\mu_0 H_{FMR}$, peak-to-peak amplitude $A_{pp}$, and peak-to-peak linewidth $W_{pp}$.  
As shown in the inset of Fig. 2(a), the change in the FMR lineshape due to current is discernible but small, such that each spectrum must be fitted to a function to extract the current-induced change in the parameters.    
The spectrum is fitted to the derivative of a generalized Lorentzian function~\cite{Stancik2008} that can accommodate small asymmetry in lineshape.   
Figures 2(b)-(d) summarize the current-induced change in the measured parameters.
Each data point is the mean of at least four measurements, and the error bar represents the standard deviation.  
Two sets of measurements were conducted with opposite magnetization alignments by applying H-fields in the +$y$ and -$y$ directions (Fig. 1).  

The most notable current-induced effect is the quadratic increase in $H_{FMR}$ (Fig. 2(b)), which arises from Joule heating that slightly reduces the saturation magnetization of FeGaB.  
In principle, it is possible to quantify a current-induced effective field,~\cite{Fan2013, Skinner2014} i.e., combination of Oersted field and field-like spin-orbit torque, from asymmetry in $H_{FMR}$ with respect to current polarity.  
We did not detect any systematic asymmetry in $H_{FMR}$, indicating that the effective field in FeGaB/Ta is below the resolution of this measurement technique.  
  
Although Joule heating is also evident in the quadratic reduction of $A_{pp}$ (Fig. 2(c)) and quadratic increase in $W_{pp}$ (Fig. 2(d)), these parameters exhibit asymmetry with respect to current polarity, as also reported in Ref.~\citenst{Ando2008a}. 
With H-field in the +$y$ direction, $A_{pp}$ is larger and $W_{pp}$ is smaller for $I_{DC}>0$, and the trend is opposite with H-field reversed.  
In quantifying the anti-damping torque, the important parameter is $W_{pp}$, which is related to the damping parameter $\alpha$ through
\begin{equation}
W_{pp} = W_{pp,inh}+\frac{4\pi f\alpha}{\sqrt{3} \gamma}, 
\end{equation}
where $W_{pp,inh}$ is the inhomogeneous linewidth broadening, $f = 9.76$ GHz and $\gamma/2\pi = 28$ GHz/T. 
While Joule heating modifies $W_{pp}$ irrespective of current polarity, the anti-damping torque changes $\alpha$ linearly with respect to current.~\cite{Ando2008a, Demidov2011, Liu2011, Pai2012, Wang2011, Wang2011a, Ganguly2014, Kasai2014}
In Fig. 2(e), we plot the linear component $\Delta W_{pp}'$ of the linewidth change, obtained by subtracting the quadratic fit background from Fig. 2(d).  
Reversing the H-field magnetizes the strip in the opposite direction and therefore reverses the polarity of the torque, represented by the slope of $\Delta W_{pp}'$ versus $I_{DC}$ (Fig. 2(e)).  
We relate the change in damping $\Delta \alpha_{eff}$, due to the anti-damping torque, to $\Delta W_{pp}'$ in Fig. 2(e) through $\Delta \alpha_{eff} = (\sqrt{3} \gamma/4 \pi f) \Delta W_{pp}'$.  

Figure 3(a) summarizes the anti-damping torque per unit charge current density $J_c$, $\Delta W_{pp}'/\Delta J_c$ (equivalently $\Delta \alpha_{eff}/\Delta J_c$), for samples with different FeGaB thicknesses.  
The trend in Fig. 3(a) is well described by the inverse thickness dependence $\sim$$1/t_F'$, where $t_F' = t_F - t_d$ is the actual ferromagnetic thickness with the effective magnetic dead layer $t_d = 0.9$ nm as elaborated below.  
This inverse magnetic thickness dependence of the current-induced torque is in agreement with a recent experimental study on CoFeB/Pt bilayers measured with quasi-static harmonic magnetization oscillation,~\cite{Fan2014} as well as with the theory that indicates the anti-damping torque from the spin Hall effect to be constant when normalized by the magnetic thickness.~\cite{Haney2013a}   
Our finding in Fig. 3(a) is thus consistent with the anti-damping torque arising from the spin Hall effect in the Ta layer.   

For practical device applications, it is of interest to estimate the threshold charge current density $J_{c,th}$ required to null damping and attain self-oscillation of magnetization.  
Assuming that Joule heating is minimized in nanoscale devices, $J_{c,th}$ can be estimated by dividing the linewidth at zero current $W_{pp,0}$ (Fig. 3(b)) by $\Delta W_{pp}'/\Delta J_{c}$ (Fig. 3(a)).  
The resulting $J_{c,th}$, plotted in Fig. 3(c), exhibits a linear decrease with respect to the magnetic layer thickness, and the range of $J_{c,th} \sim 10^{11}-10^{12}$ A/m$^{2}$ is consistent with recent reports of spin-Hall-driven self-oscillation of magnetization.\cite{Liu2012a, Demidov2012, Liu2013, Zholud2014}  
From Fig. 3(c), one might conclude that a thinner magnetic layer allows for a more efficient spin-Hall oscillator. 
However, as the magnetic layer thickness is reduced, the damping increases due to increased contributions from interfacial defects and spin pumping, and the resonance signal deteriorates.    
For example, we could not detect any spin-Hall anti-damping torque in FeGaB/Ta strips with an even thinner FeGaB layer ($t_F = 1.5$ nm), because of poor FMR signal with a very broad linewidth $W_{pp,0} \approx 20$ mT.  

We now quantify the effective spin Hall angle $\theta^{eff}_{SH}$ from the DC-induced linewidth change, assuming a very large spin-mixing conductivity $G_{\uparrow\downarrow}$ at the FeGaB-Ta interface and $t_{Ta} \gg \lambda_{Ta}$.  
For the case where the external field is in the film plane and orthogonal to the charge current direction:~\cite{Petit2007, Liu2011}
\begin{equation}
|\theta^{eff}_{SH}| = \left( H_{ext}+\frac{M_{eff}}{2} \right) \mu_0 M_s' t_F' \frac{2|e|}{\hbar} \left| \frac{\Delta\alpha_{eff}}{\Delta J_c} \right|.
\end{equation}
Here, $M_s'$ and $t_F'$ are the actual saturation magnetization and thickness, respectively, of the ferromagnetic layer as quantified below.  
$H_{ext}$ is the field at resonance taken to be $H_{FMR,0}$ as summarized in Fig. 4(a).  
$M_{eff}$ is the effective saturation magnetization incorporating out-of-plane magnetic anisotropy, obtained from the Kittel equation $f = (\gamma/2\pi)\mu_0\sqrt{H_{FMR,0}(H_{FMR,0}+M_{eff})}$.  

Figure 4(b) shows, along with $M_{eff}$, the saturation magnetization $M_{s}$ of unpatterned films measured with vibrating sample magnetometry normalized by the nominal ferromagnetic volume.  
$M_{eff}$ and $M_{s}$ are in close agreement, indicating negligible perpendicular magnetic anisotropy in these FeGaB/Ta films.  
The reduction of $M_s$ with decreasing $t_F$ implies the existence of a magnetic dead layer of thickness $t_d$, such that the actual ferromagnetic thickness is $t_F' = t_{F} - t_d$.  
The saturation magnetization normalized with respect to the film area (inset Fig. 4(b)) vanishes at $t_F = 0.9$ nm, corresponding to the estimated $t_d$.  
The slope gives $\mu_0 M_s' = 1.4$ T, in agreement with the saturation magnetization of thick Fe$_7$Ga$_2$B$_1$ films.~\cite{Lou2007}   From Eq. 2, we obtain $|\theta^{eff}_{SH}|= 0.08-0.09$ independent of the ferromagnetic thickness (Fig. 4(c)), consistent with the Ta layer acting as the source of the spin-Hall anti-damping torque.\cite{footnoteUncorrectedMs}     
This range is similar to the higher end of reported effective spin Hall angles in various ferro(ferri)magnet/Ta bilayers, $|\theta^{eff}_{SH}| \approx 0.1$, estimated from magnetization switching,~\cite{Liu2012} domain-wall displacement,~\cite{Emori2013a} and electrical detection of spin pumping.~\cite{Wang2014}  

Taking into account finite $G_{\uparrow\downarrow}$ and assuming $t_{Ta} \gg \lambda_{Ta}$, we find from a model of spin-diffusive transport, similar to Refs. \citenst{Haney2013a} and \citenst{Boone2013}, that the intrinsic spin Hall angle $\theta_{SH}$ of Ta is given by $\theta_{SH}=\theta^{eff}_{SH}(2G_{\uparrow\downarrow}+\sigma_{Ta}/\lambda_{Ta})/2G_{\uparrow\downarrow}$, where $\sigma_{Ta}$ is the conductivity of Ta.  
This implies that intrinsic $\theta_{SH}$ of Ta may be larger than $\theta^{eff}_{SH}$ estimated from our results, and that extracting $\theta_{SH}$ requires careful quantification of  $G_{\uparrow\downarrow}$ in FeGaB/Ta from spin-pumping-induced increase in $\alpha$.\cite{Boone2013}  
Nevertheless, our simple cavity-based measurement technique allows for quantifying $\theta^{eff}_{SH}$ as the lower-bound for $\theta_{SH}$, and $\theta^{eff}_{SH}$ is a useful figure of merit indicating how much spin current is injected into the ferromagnet for a given input charge current density.   
Moreover, the spin-Hall anti-damping torque may be enhanced with engineering of the ferromagnet-Ta interface (increasing $G_{\uparrow\downarrow}$), for instance by tuning the interfacial magnetic dead layer.  

We attempted as an independent check to quantify the spin-Hall anti-damping torque in 10-$\mu$m wide, 50-$\mu$m long FeGaB/Ta strips from linewidth modification in ST-FMR spectra.~\cite{Liu2011, Pai2012, Ganguly2014, Kasai2014}  
However, the signal-to-noise ratio of the spectra was not sufficient to extract a systematic change of resonance linewidth with respect to DC current, because of the high resistivity and low anisotropic magnetoresistance ($<0.03\%$) of the FeGaB/Ta devices.   
This difficulty with ST-FMR highlights the advantage of using a cavity-based resonance spectrometer, which enables detection of small changes in the resonance linewidth without relying on a magnetoresistance signal.  
The cavity-based technique is therefore applicable to any material systems with sufficiently narrow resonance linewidth, including electrically insulating thin films of yttrium iron garnet interfaced with spin-Hall metals.      

In summary, we have quantified the spin-Hall anti-damping torque in FeGaB/Ta bilayers using a sensitive cavity-based resonance spectrometer.  
Systematic linewidth changes of $\sim$10 $\mu$T induced by small current densities $\sim$10$^9$ A/m$^2$ are detected.   
The extracted effective spin Hall angle of 0.08-0.09 is independent of the ferromagnetic layer thickness, consistent with the simple mechanism where the Ta layer acts as the source of the anti-damping torque.  
This technique based on a resonance spectrometer is generally suitable for investigating the anti-damping torque in other material systems, even those with vanishingly small magnetoresistance.  
\\
\\
This work was supported by the Air Force Research Laboratory.  Lithography was performed in the George J. Kostas Nanoscale Technology and Manufacturing Research Center, and x-ray reflectivity was performed in the MIT Center for Materials Science and Engineering.

\newpage
\begin{figure}[h]
\includegraphics[]{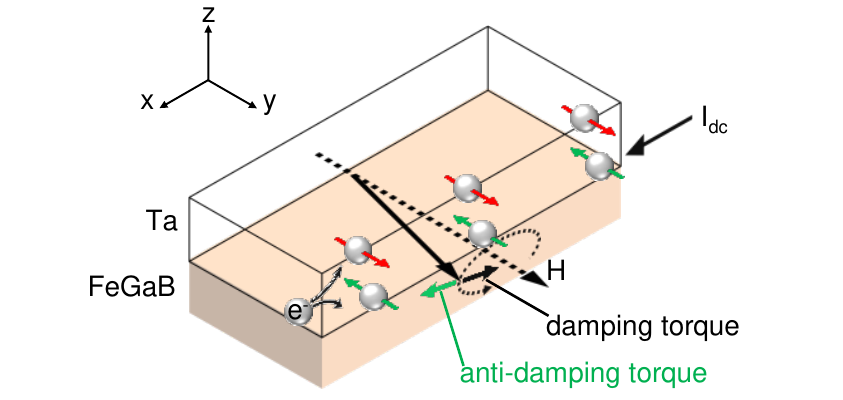}
\caption{\label{fig:cartoon} Schematic of the anti-damping torque generated by the spin Hall effect in Ta.} 
\end{figure}

\begin{figure}[h]
\includegraphics[]{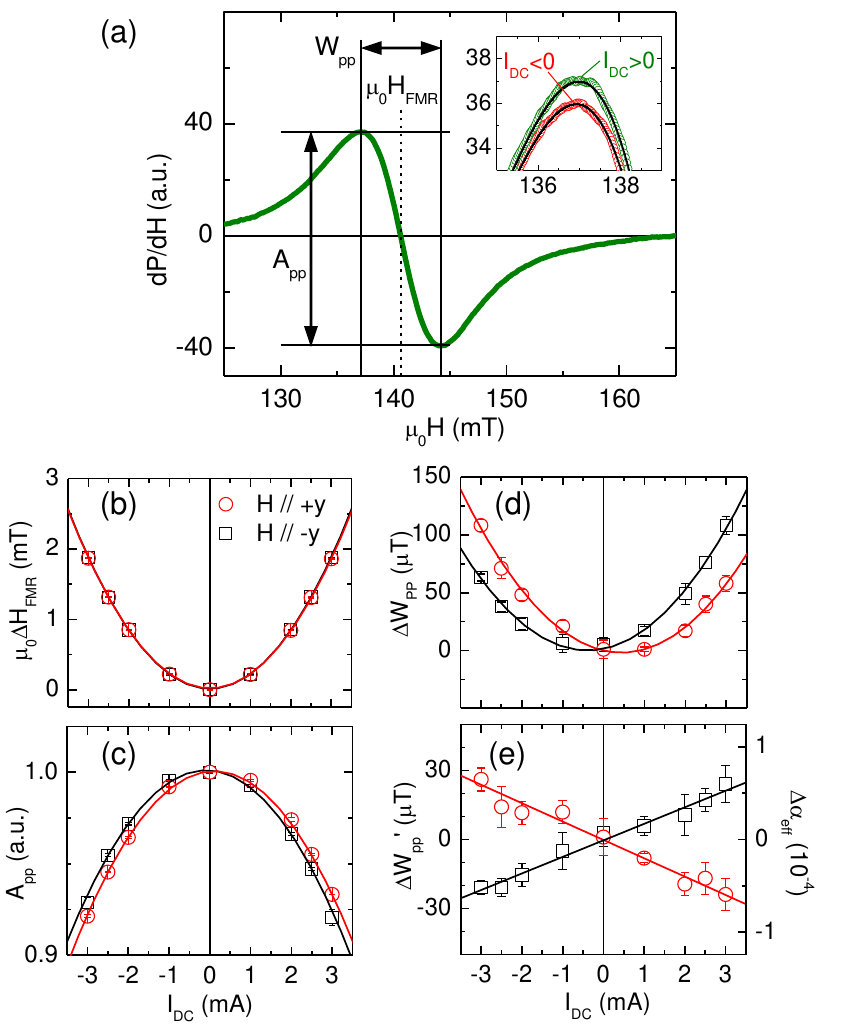}
\caption{\label{fig:FMRexample} (a) Definitions of resonance field $H_{FMR}$, peak-to-peak amplitude $A_{pp}$, and peak-to-peak linewidth $W_{pp}$.  Inset: Modification of FMR spectrum at DC currents of opposite polarity.  The spectra are from a 100-$\mu$m-wide, $t_F =$ 2.2 nm sample at $|I_{DC}| = 3.2$ mA.  (b-d) Current-induced changes in (b) $H_{FMR}$, (c) normalized $A_{pp}$, (d) $W_{pp}$, and (d) $W_{pp}$ with quadratic background subtracted, in a 100-$\mu$m wide $t_F =$ 3.4 nm strip.} 
\end{figure}

\begin{figure}
\includegraphics[]{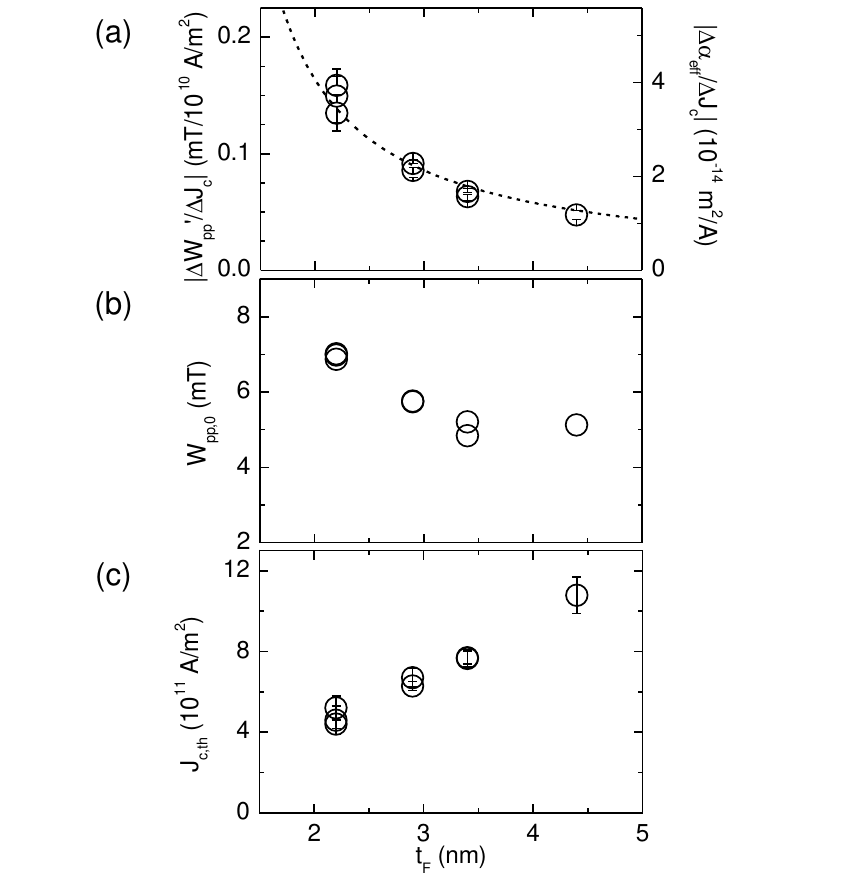}
\caption{\label{fig:thickness} Dependence on FeGaB thickness of (a) current-induced linear modification in linewidth $W_{pp}$ (effective damping $\alpha_{eff}$), (b) linewidth $W_{pp,0}$ at zero current, and (c) estimated threshold charge current density $J_{c,th}$ required to null damping. The dotted curve in (a) shows the fit to $\sim$$1/(t_F-t_d)$ with $t_d = 0.9$ nm.} 
\end{figure}

\begin{figure}
\includegraphics[]{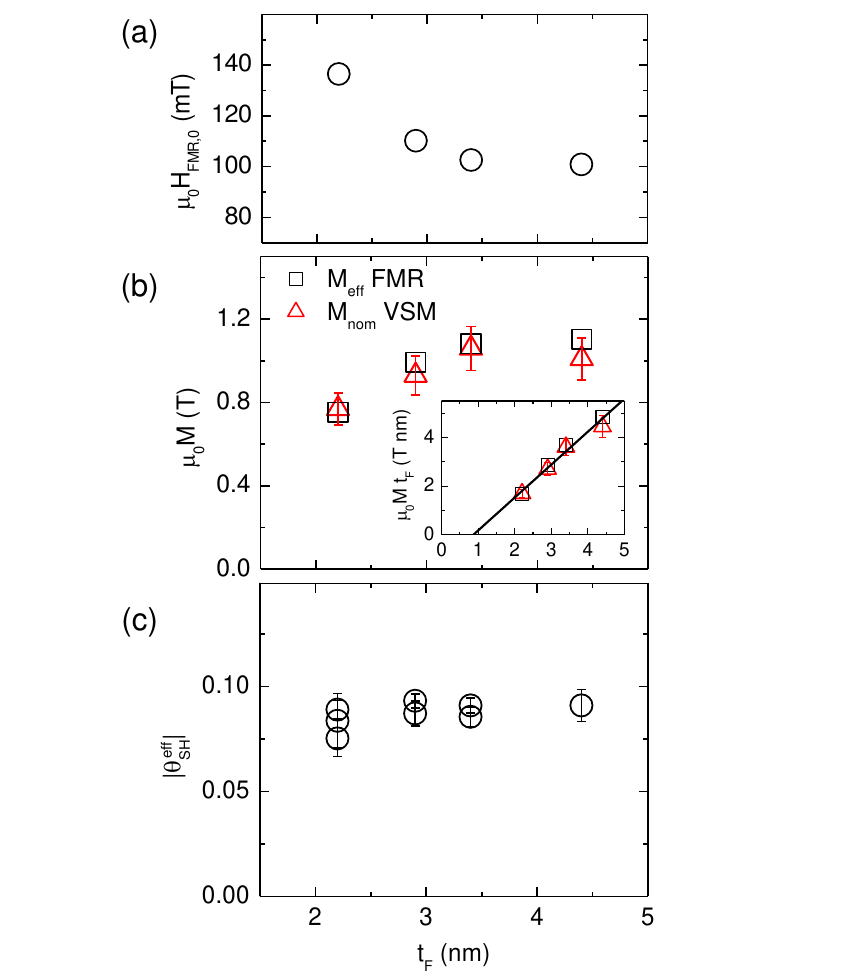}
\caption{\label{fig:STFMR} (a) Resonance field $H_{FMR,0}$ at zero charge current.  (b) Effective magnetization $M_{eff}$ obtained from $H_{FMR,0}$ and saturation magnetization $M_{nom}$ measured with vibrating sample magnetometery normalized by nominal ferromagnetic volume. Inset: Magnetization normalized by film area.  (c) Effective spin Hall angle $|\theta^{eff}_{SH}|$.  All data are plotted with respect to the nominal FeGaB layer thickness $t_F$. 
} 
\end{figure}


\begin{thebibliography}{10}
\bibitem{footnoteUncorrectedMs}
Using the uncorrected $M_s$ and $t_F$ in Eq. 2 gives a nearly identical range
  of $|\theta^{eff}_{SH}|$, except for $t_F = 2.2$ nm where the resulting
  $|\theta^{eff}_{SH}|$ is somewhat lower at 0.06-0.07.
\end{thebibliography}
\end{document}